# 垃圾焚烧电厂-P2G协同运行下考虑余热利用的综合能源系统低碳优化调度

王利猛，王　硕，王　娜，马语泽，李　扬

（现代电力系统仿真控制与绿色电能新技术教育部重点实验室 (东北电力大学)，吉林 132012）

**摘　要**：为提高能源利用率和降低碳排放，提出一种含余热回收利用的垃圾焚烧电厂-电转气(power-to-gas，P2G)的综合能源系统经济运行策略。首先考虑垃圾焚烧电厂-P2G的协同运行，引入细化电转气两阶段运行过程，在传统电转气的基础上加入氢燃料电池以减少能量阶梯损耗，并将甲烷化反应热回收利用；其次为提高垃圾焚烧能源利用效率，考虑加装含水源热泵的余热回收装置，通过消耗部分电能来回收烟气余热以及消纳风电，并增设$CO_2$分离装置，结合P2G将回收的$CO_2$合成$CH_4$，实现碳的循环利用。最后基于阶梯式碳交易机制构建以系统运行成本最小为目标的电-热综合能源系统优化模型，利用GUROBI建模优化引擎对此模型进行求解，结果验证所建模型的有效性。

## 0　引　言

以双碳为目标的低碳发展方式对解决新能源和传统发电机组之间的矛盾，实现多能源之间的相互耦合具有重要意义[1-2]。垃圾焚烧发电技术不仅有效解决了城市生活垃圾，且提高了能源可持续性发展的可能[3-4]。垃圾焚烧发电是有效处理生活废物的方式之一，文献[5]构建了以垃圾焚烧电厂供能效率、碳排放、经济效益三者间的多目标优化模型，平衡三者之间的关系。文献[6]考虑了含风电与垃圾焚烧电厂的 IES（integrated energy system，IES）优化调度模型，协调源荷两侧，不仅适应了风电的反调峰特性，提高新能源利用率，且提高了系统整体经济性。文献[7]利用电储能提供微电网运行所需的旋转备用以平抑新能源发电的波动性。

然而，鲜有文献将系统中产生的高温烟气的余热进行详细研究。余热利用技术因其节能性，现已在工业领域受到广泛重视[8]。对低温余热进行高效利用，不仅可提高机组能效，且对中国实现双碳目标具有重要意义[9-10]。文献[11]提出传统接触式换热器与压缩式热泵耦合的烟气冷凝余热回收系统，实现了余热的高效回收。文献[12]引入含余热回收的电转气设备，优化电-热-气耦合关系来消纳新能源并提高碳减排水平。

电转气技术将风电、光伏转换成甲烷供给燃汽机组或天然气市场，实现新能源的转换和消纳，达到电负荷转移以及削峰填谷的目的[13-15]。文献[16]提出 CHP 机组和 P2G 设备协同运行的优化调度模型以提高新能源消纳水平。然而，鲜有文献考虑 P2G 中甲烷化反应阶段所产生的余热，若将其回收利用，可缓解 IES 内供热设备的压力，提升能源利用效率、节省运行成本、降低碳交易成本，延伸 P2G 消纳风电的价值，且目前对 P2G 原料的来源及成本进行的研究较少，垃圾焚烧电厂烟气经固有的净化装置处理后烟气成分简单，$CO_2$ 分离相对碳捕集能耗较低，可为 P2G 设备中的甲烷化反应装置提供 $CO_2$ 原料。碳交易机制是实现低碳减排目标的有效方式之一，近年来提出的阶梯型碳交易机制相比于传统型碳交易机制，其对碳排放量的控制更为严格[17-18]。文献[19]建立了含阶梯式碳交易机制的冷、热、电负荷的综合需求响应模型。文献[20]考虑了阶梯式碳交易机制下用户参与需求响应的混合整数非线性优化模型。结果表明，考虑阶梯式碳交易机制不仅能达到削峰填谷的目的，且可降低碳排放和系统运行总成本。

本文在上述研究的基础上，提出基于阶梯式碳交易机制并考虑余热回收利用的垃圾焚烧电厂-P2G 联合运行的 IES 低碳经济运行模型，模型中考虑外部购电、阶梯式碳交易、弃风弃光等成本，以 IES 低碳运行成本最小为目标，并利用 GUROBI 建模优化引擎进行求解。

## 1　考虑余热利用的垃圾焚烧电厂-P2G 的 IES 低碳经济运行系统

IES 结合各类电-热-气设备进行多种能源之间的合理调配来响应不同种类负荷需求。建立在市区附近的垃圾焚烧电厂相较于偏远的火电厂，更具地理位置优势、更有利于城市的供电供暖服务。本文在传统垃圾焚烧电厂模型基础上，细化考虑电转气设备中氢能的高效利用及甲烷化反应的余热回收，并回收垃圾焚烧经净化后排放的高温烟气，且引入阶梯式碳交易机制，所构建的具体模型框架如图 1 所示。所建立的综合能源系统主要包括 3 个部分：能量供给侧部分包括电网、天然气市场、

收稿日期：2023-08-06

基金项目：吉林省自然科学基金（YDZJ202101ZYTS149）

通信作者：王利猛（1971—）男，博士、副教授、硕士生导师，主要从事电力系统监测与控制方面的研究。wlm_28@163.com

风电机组、垃圾焚烧电厂；耦合设备部分包括电解槽、甲烷化反应设备、燃气锅炉机组等能量转换设备；终端用能部分包括电-热-气三类负荷。

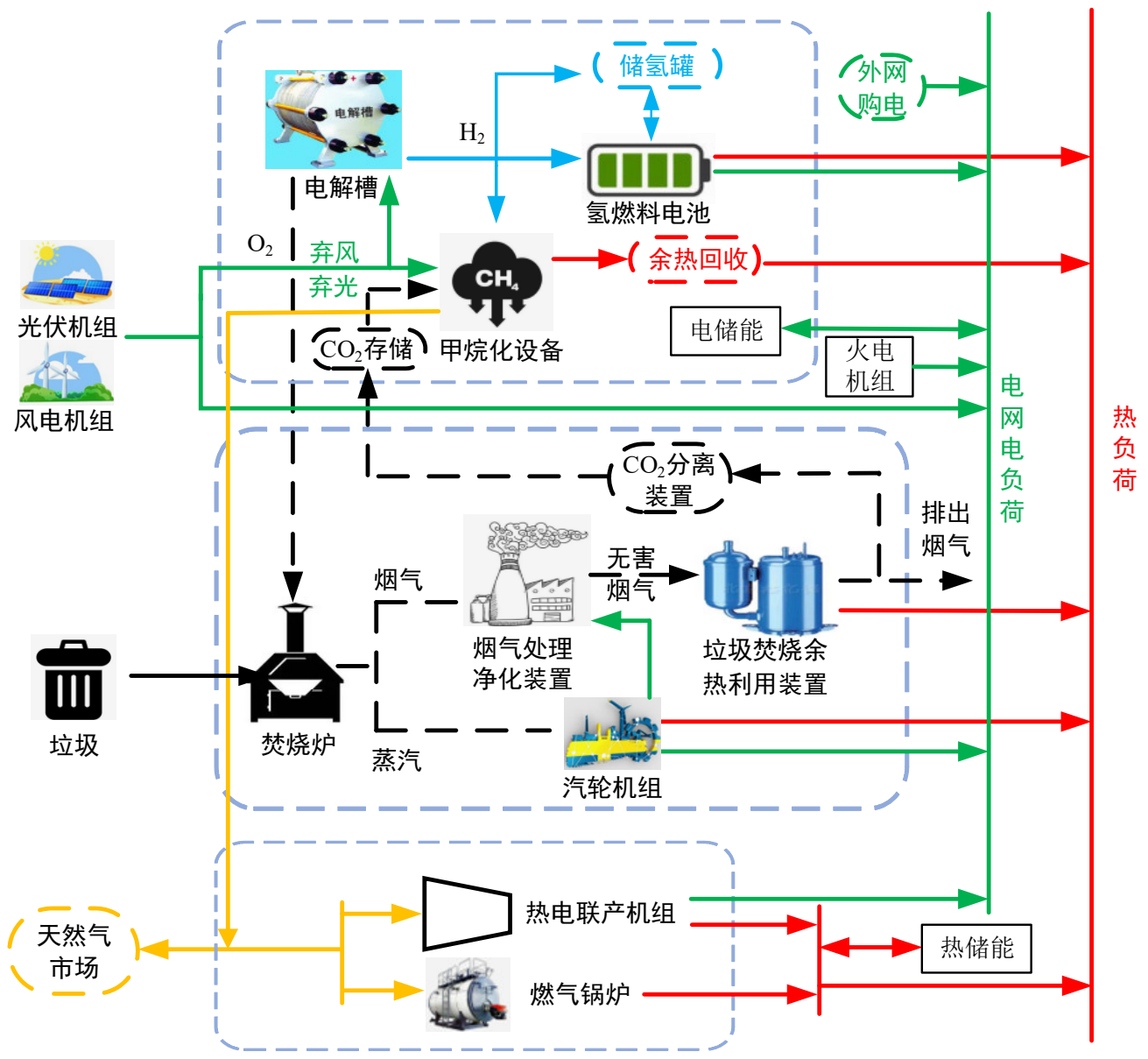

图 1　含垃圾焚烧电厂及细化电转气设备的 IES

Fig. 1　IES of waste incineration plant and hydrogen production

## 2　垃圾焚烧电厂协同电转气低碳运行

### 2.1　垃圾焚烧-烟气处理-余热回收系统

垃圾焚烧-烟气处理-余热回收系统由焚烧炉、烟气净化处理装置、余热利用装置等组成。首先将城市垃圾送入储存仓内，进行静置、脱水、干燥等预处理过程，然后将其送入焚烧炉中烧释放内部化学能，产生的水蒸气送入汽轮机组生成电能和热能。

#### 2.1.1 垃圾焚烧电厂出力及爬坡约束

由于垃圾焚烧电厂日处理垃圾量恒定，因此其日总出力为一定值。

$$\begin{cases} W^{\mathrm{WI}} = \sum_{t=1}^{T} P_t^{\mathrm{WI}} \\ P^{\mathrm{WI,\,min}} \le P_t^{\mathrm{WI}} \le P^{\mathrm{WI,\,max}} \\ \left| P_{t+1}^{\mathrm{WI}} - P_t^{\mathrm{WI}} \right| \le \Delta P^{\mathrm{WI}} \end{cases} \tag{1}$$

式中：$W^{\mathrm{WI}}$——垃圾焚烧电厂日恒定总出力，MWh；$P^{\mathrm{WI,max}}$、$P^{\mathrm{WI,min}}$——$t$ 时刻垃圾焚烧电厂出力最大值、最小值，MW。

#### 2.1.2 烟气处理与余热利用

垃圾在焚烧过程中排出大量含硫化物、二噁英、颗粒物等有害物质的烟气，需送入烟气净化装置中处理合格后排放。经处理后的烟气温度一般约为 160 ℃，含有大量潜热的水蒸气，若直接排出会浪费大量能源。日恒处理 500 t 垃圾的焚烧炉产生的烟气中可回收约 300 MWh 的余热量，且 1 t 垃圾约产生 0.3 t 的 $CO_2$，随着碳交易市场的发展，若排放超过限额的 $CO_2$ 将有惩罚成本。为了尽可能多的回收烟气中的余热和降低 $CO_2$ 的排放，本文在传统烟气处理净化装置的基础上加装如图 2 所示的 $CO_2$ 分离装置与余热回收利用装置。

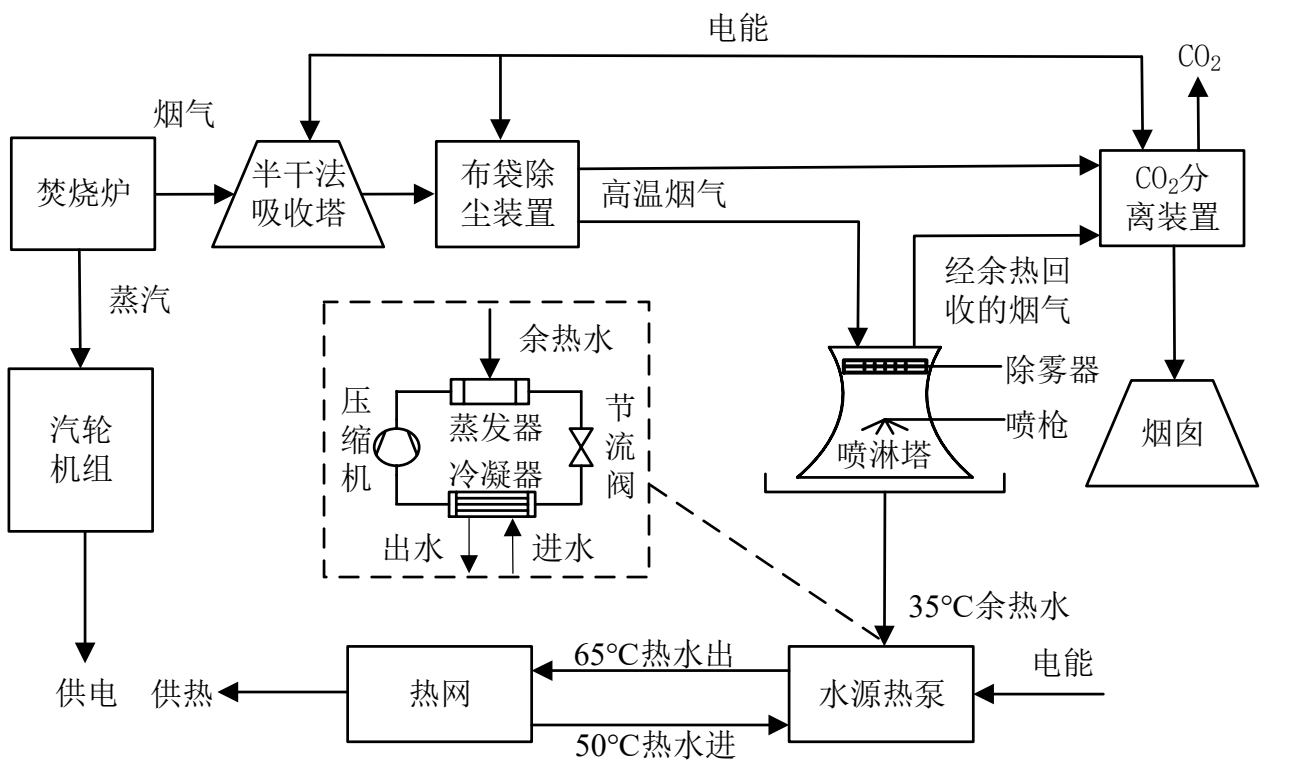

图 2　$CO_2$ 分离装置与余热回收利用装置

Fig. 2　$CO_2$ separation device and waste heat recovery and utilization device

焚烧炉排出的烟气首先经过烟气净化处理装置（吸收塔、布袋除尘装置）完成对烟气排放出的有害物质的吸附拦截。喷淋塔顶端为除雾区，防止水蒸气和雾气随净化处理后的高温烟气排出，塔内设置3~4层喷枪喷出循环雾化水与高温烟气混合交换热量，使烟气温度从160 ℃下降到45 ℃放出显热，烟气包含的水蒸气冷凝放出隐热，从而将气体中的热量交换到水中。换热完成后烟气冷却水温度可达35 ℃，但该温度并未达到北方城市规定的集中供暖水温，因此进一步采用水源热泵设备，以低温热源（冷却后的余热水）为驱动热源，经过蒸发、压缩、冷凝等步骤，仅需消耗少量电能即可达到从低温向高温输送热能的目的。喷淋塔出口烟气在引风机作用下进入$CO_2$分离装置，经上述处理后，烟气中剩余成分主要为$CO_2$、$O_2$和$N_2$，因此仅需对其进行简单的分离处理，并将分离的$CO_2$体积量换算成质量，即可向P2G中的甲烷化反应设备输送$CO_2$作为原料，该方法只需用较小的能耗即可等效于$CO_2$捕集装置的作用。烟气分离与余热回收利用装置模型如下文所述。

$t$ 时刻烟气分离装置分离 $CO_2$ 的耗电量为 $P_{\mathrm{Cf},t}$，即：

$$P_{\mathrm{Cf},t} = \lambda \cdot V_{\mathrm{C},t} \tag{2}$$

$$V_{\mathrm{C},t} = \gamma \cdot \beta \cdot V_{0,t} \tag{3}$$

式中：$\lambda$——分离单位 $CO_2$ 的耗电量，$m^3$/MWh；$V_{\mathrm{C},t}$——$t$ 时刻分离出的 $CO_2$ 的体积，$m^3$；$\gamma$——烟气中 $CO_2$ 的占比；$\beta$——$CO_2$ 分离率。

余热回收装置回收的余热量为：

$$Q_1 = \lambda_1 \cdot Q_{\mathrm{Y}} \tag{4}$$

式中：$\lambda_1$——喷淋塔能效利用率；$Q_{\mathrm{Y}}$——烟气中所含能量，MWh。

中国北方城市供热二次网供水温度一般为55~65 ℃，回水温度一般为 50 ℃，水源热泵能效系数COP 进水及供水侧水温关系[21]如图 3 所示。根据水源热泵设备原理可知，余热水源的进水温度与热泵换热效率成正相关，因此提高余热水温度可提高换热效率。

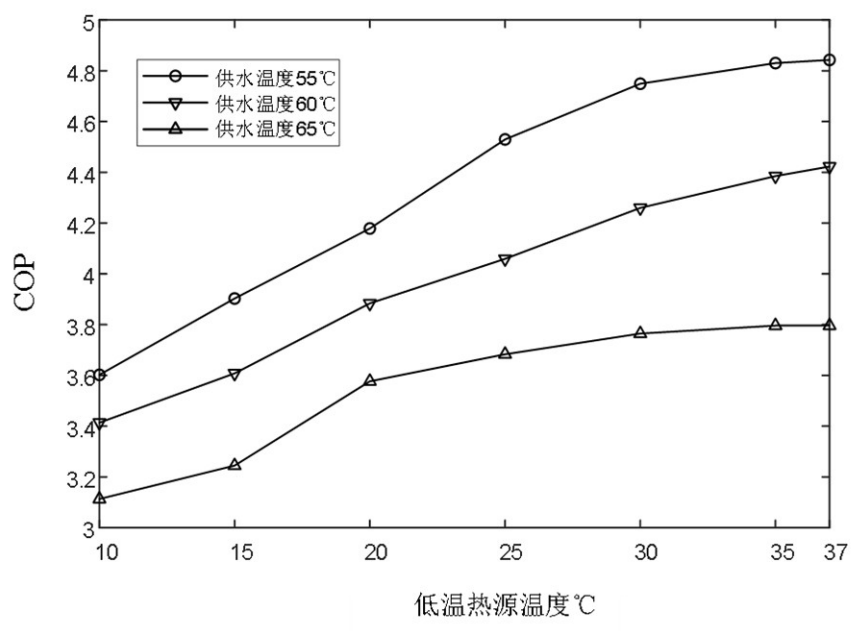

图 3　COP 与低温热源温度关系

Fig. 3　Relationship between COP and temperature of low-temperature heat sources

水源热泵能效函数为：

$$P_{p,t}\cdot\varepsilon_{COP}=Q_{p,t} \tag{5}$$

式中：$P_{p,t}$——$t$ 时刻水源热泵耗电功率，MW；$Q_{p,t}$——$t$ 时刻水源热泵出口热量，MW；$\varepsilon_{COP}$——水源热泵能效系数。

水源热泵约束为：

$$0\le P_{p,t}\le P_{P}^{max} \tag{6}$$

式中：$P_{P}^{max}$——水源热泵设备耗电功率的最大值，MW。

拟合该设备的低温热源温度-COP 关系得到：

$$\varepsilon_{COP}=f\left(T_{temp}\right) \tag{7}$$

式中：$f\left(T_{temp}\right)$——低温热源余热水温度，℃。

## 2.2　细化电转气设备及运行过程

氢能作为清洁、高效的能源，受到广泛关注。含电制氢的细化电转气运行过程如图 4 所示。电解槽（electrolyzer，EL）将电能转换为 $H_2$ 和 $O_2$，将 $H_2$ 输送至甲烷反应器（methane reactor，MR）与 $CO_2$ 合成天然气，氢能还可直接储存于氢燃料电池（hydrogen fuel cell，HFC）或储氢罐中，以便于直接转换为电能和热能使用，相比于转化为天然气少了一个能量转化环节，可减少能量的梯级损耗。富余天然气可供给热电联产机组、燃气锅炉或天然气市场。已知每消耗 1 kWh 的电能，可回收的甲烷化反应热为 0.1188 kWh。

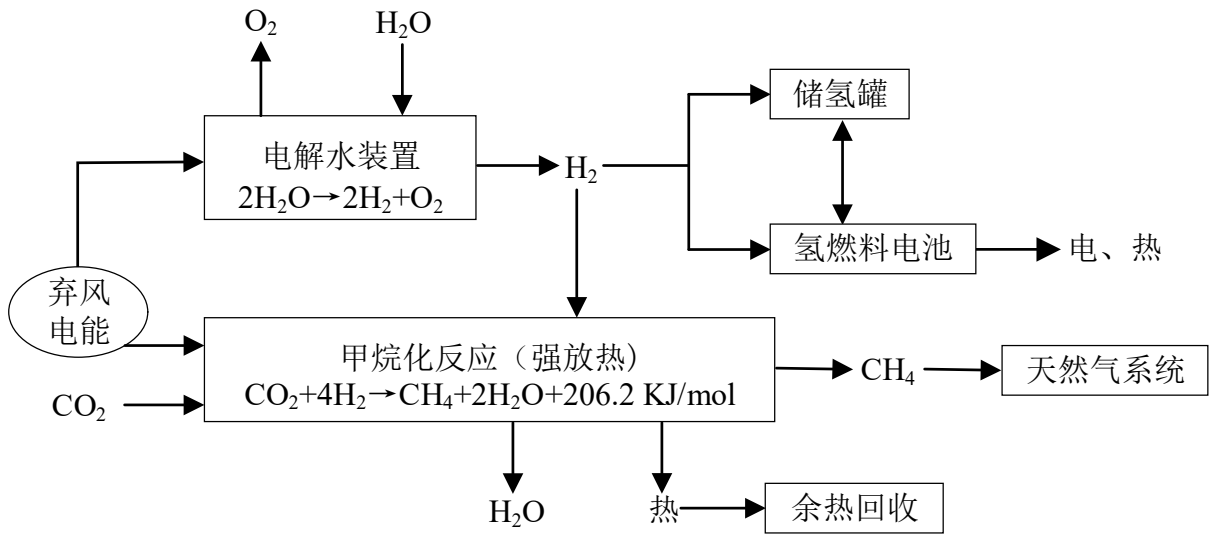

图4　细化电转气及电制氢流程图

Fig. 4　Refining the flow chart of electrolytic hydrogen

1）EL 设备

$$\begin{cases}P_{EL,H_2}(t)=\eta_{EL}P_{e,EL}(t)\\P_{e,EL}^{min}\le P_{e,EL}\le P_{e,EL}^{max}\\\Delta P_{e,EL}^{min}\le P_{e,EL}(t+1)-P_{e,EL}(t)\le\Delta P_{e,EL}^{max}\end{cases} \tag{8}$$

式中：$P_{EL,H_2}(t)$——$t$ 时刻 EL 输出的氢功率，MW；$P_{e,EL}(t)$——$t$ 时刻输入 EL 的电功率，MW；$P_{e,EL}^{max}$、$P_{e,EL}^{min}$——输入 EL 电功率的最大、最小值，MW；$\Delta P_{e,EL}^{max}$、$\Delta P_{e,EL}^{min}$——EL 的爬坡功率最大、最小值，MW；$\eta_{EL}$——EL 过程的能量转换效率。

2）MR 设备

$$\begin{cases}P_{MR,g}(t)=\eta_{MR}P_{e,MR}(t)\\P_{H_2,MR}^{min}\le P_{H_2,MR}\le P_{H_2,MR}^{max}\\\Delta P_{H_2,MR}^{min}\le P_{H_2,MR}(t+1)-P_{H_2,MR}(t)\le\Delta P_{H_2,MR}^{max}\end{cases} \tag{9}$$

式中：$P_{MR,g}(t)$——$t$ 时刻 MR 输出天然气功率，MW；$P_{H_2,MR}(t)$——$t$ 时刻 MR 的输入氢功率，MW；$P_{H_2,MR}^{max}$、$P_{H_2,MR}^{min}$——MR 的输入氢功率的最大、最小值，MW；$\Delta P_{H_2,MR}^{max}$、$\Delta P_{H_2,MR}^{min}$——MR 爬坡功率的最大、最小值，MW；$\eta_{MR}$——MR 过程的能量转换效率。

3）HFC 设备

$$\begin{cases}P_{HFC,e}(t)=\beta_{HFC}^{e}P_{H_2,HFC}\\P_{HFC,h}(t)=\beta_{HFC}^{h}P_{H_2,HFC}\\P_{H_2,HFC}^{min}\le P_{H_2,HFC}(t)\le P_{H_2,HFC}^{max}\\\Delta P_{H_2,HFC}^{min}\le P_{H_2,HFC}(t+1)-P_{H_2,HFC}(t)\le\Delta P_{H_2,HFC}^{max}\\P_{HFC,h}(t)/P_{HFC,e}(t)=\lambda_{HFC}\end{cases} \tag{10}$$

式中：$P_{H_2,HFC}$——$t$ 时刻 HFC 的输入氢功率，MW；$P_{HFC,e}(t)$、$P_{HFC,h}(t)$——$t$ 时刻 HFC 的输出电、热功率，MW；$P_{H_2,HFC}^{max}$、$P_{H_2,HFC}^{min}$——HFC 的输入氢能最大、最小值，MW；$\Delta P_{H_2,HFC}^{max}$、$\Delta P_{H_2,HFC}^{min}$——HFC 爬坡功率的最大、最小值，MW；$\lambda_{HFC}$——HFC 的热电比；$\beta_{HFC}^{e}$、$\beta_{HFC}^{h}$——HFC 转化成电、热功率的效率。

## 2.3　CHP机组和燃气锅炉模型

$$\begin{cases}P_{CHP,e}(t)=\eta_{CHP}^{e}P_{g,CHP}(t)\\P_{CHP,h}(t)=\eta_{CHP}^{h}P_{g,CHP}(t)\\P_{g,CHP}^{min}\le P_{g,CHP}(t)\le P_{g,CHP}^{max}\\\Delta P_{g,CHP}^{min}\le P_{g,CHP}(t+1)-P_{g,CHP}(t)\le\Delta P_{g,CHP}^{max}\\k_{CHP}^{min}\le P_{CHP,h}(t)/P_{CHP,e}(t)\le k_{CHP}^{max}\end{cases} \tag{11}$$

$$P_{GB}(t)=V_{GB}H_{g}\eta_{GB} \tag{12}$$

式中：$P_{CHP,e}(t)$、$P_{CHP,h}(t)$、$P_{g,CHP}(t)$——$t$ 时段 CHP 输出的电、热及天然气功率，MW；$\eta_{CHP}^{e}$、$\eta_{CHP}^{h}$——天然气在机组内转换为电、热能的效率；$P_{g,CHP}^{max}$、$P_{g,CHP}^{min}$——输入 CHP 的天然气功率的最大、最小值，MW；$\Delta P_{g,CHP}^{max}$、 $\Delta P_{\mathrm{g,CHP}}^{\min}$——CHP 的爬坡上、下限，MW；$k_{\mathrm{CHP}}^{\max}$、$k_{\mathrm{CHP}}^{\min}$——CHP 的热电比上、下限；$P_{\mathrm{GB}}(t)$——$t$ 时段 GB 机组输出热功率，MW；$V_{\mathrm{GB}}$——$t$ 时段 GB 机组消耗的天然气量，$\mathrm{m}^3$；$\eta_{\mathrm{GB}}$——GB 能量转化效率。

### 2.4 奖惩阶梯式碳交易成本计算模型

在碳交易机制下，碳排放量是可以自由买卖无偿交易的，无偿碳排放额分配：

$$\begin{cases} E_{总} = E_{\mathrm{PN}} + E_{\mathrm{CHP}} + E_{\mathrm{GB}} \\ E_{\mathrm{GB}} = \beta_{\mathrm{h}} \sum_{t=1}^{T} H_{\mathrm{GB}}(t)\Delta t \\ E_{\mathrm{PN}} = \beta_{\mathrm{e}} \sum_{t=1}^{T} P_{\mathrm{PN}}(t)\Delta t \\ E_{\mathrm{CHP}} = \beta_{\mathrm{h}} (\sum_{t=1}^{T} H_{\mathrm{CHP}}(t)\Delta t + \phi_{\mathrm{e,h}} \sum_{t=1}^{T} P_{\mathrm{CHP}}(t)\Delta t) \end{cases} \tag{13}$$

式中：$E_{总}$、$E_{\mathrm{CHP}}$、$E_{\mathrm{PN}}$、$E_{\mathrm{GB}}$——综合能源系统、CHP 系统、外部购电，GB 的无偿碳排放配额，t；$\beta_{\mathrm{e}}$、$\beta_{\mathrm{h}}$——单位电量、热量的无偿碳排放系数；$P_{\mathrm{PN}}(t)$——$t$ 时刻系统的购电功率，MW；$H_{\mathrm{GB}}(t)$——$t$ 时刻燃气锅炉的输出热功率，MW；$P_{\mathrm{CHP}}(t)$、$H_{\mathrm{CHP}}(t)$——$t$ 时刻热电联产的电能、热能的输出功率，MW；$\phi_{\mathrm{e,h}}$——CHP 设备发电量的等效热量换算系数。

阶梯式碳排放交易成本 $f_{\mathrm{CO_2}}^{\mathrm{price}}$ 计算：

$$\begin{cases} -cd - c(1+\alpha)(-d-\Delta E), & \Delta E < -d \\ c\Delta E, & -d \le \Delta E < 0 \\ c\Delta E, & 0 \le \Delta E < d \\ cd + c(1+\alpha)(\Delta E - d), & d \le \Delta E < 2d \\ c(2+\alpha)d + c(1+2\alpha)(\Delta E - 2d), & 2d \le \Delta E < 3d \\ c(3+3\alpha)d + c(1+3\alpha)(\Delta E - 3d), & 3d \le \Delta E < 4d \\ c(4+6\alpha)d + c(1+4\alpha)(\Delta E - 4d), & 4d \le \Delta E < 5d \\ c(5+10\alpha)d + c(1+5\alpha)(\Delta E - 5d), & 5d \le \Delta E < 6d \\ c(6+15\alpha)d + c(1+6\alpha)(\Delta E - 6d), & 6d \le \Delta E \end{cases} \tag{14}$$

式中：$\alpha$——碳交易价格增长系数；$c$——碳交易基价，\$/t；$d$——碳交易价格区间长度，t；$\Delta E$——综合能源系统实际碳排放量与无偿碳排放配额的差值，t。

由于 P2G 过程会消纳 $\mathrm{CO_2}$，因此 IES 的实际碳排放为：

$$\begin{cases} E_{总} = E_{\mathrm{PN}} + E_{\mathrm{CHP}} + E_{\mathrm{GB}} - E_{\mathrm{P2G}} - E_{\mathrm{MR}} \\ E_{\mathrm{P2G}} = \beta_{\mathrm{g}} \sum_{t=1}^{T} P_{\mathrm{P2G}}(t)\Delta t \\ E_{\mathrm{MR}} = \beta_{\mathrm{R}} \sum_{t=1}^{T} P_{\mathrm{MR}}(t)\Delta t \end{cases} \tag{15}$$

式中：$E_{\mathrm{GB}}$、$E_{\mathrm{MR}}$——GB、MR 的实际碳排放量，t；$\beta_{\mathrm{g}}$、$\beta_{\mathrm{R}}$——取值不同的 P2G、MR 设备的碳排放系数；$E_{\mathrm{P2G}}$、$E_{\mathrm{MR}}$——P2G、MR 设备捕获的 $\mathrm{CO_2}$ 量，t；$P_{\mathrm{P2G}}(t)$、$P_{\mathrm{MR}}(t)$——$t$ 时刻 P2G、MR 设备的输入功率，MW。

## 3 基于阶梯式碳交易机制的 IES 优化运行模型

### 3.1 目标函数

本文以 IES 低碳经济运行成本最小为目标，构建目标函数：

$$\begin{aligned} F = \min(& f_{\mathrm{buy}}^{\mathrm{price}} + f_{\mathrm{CO_2}}^{\mathrm{price}} + f_{\mathrm{WI}}^{\mathrm{price}} + f_{\mathrm{PH}}^{\mathrm{price}} + \\ & f_{\mathrm{P2G}}^{\mathrm{price}} + f_{\mathrm{Q,cut}}^{\mathrm{price}} + f_{\mathrm{W}}^{\mathrm{price}} + f_{\mathrm{G}}^{\mathrm{price}}) \end{aligned} \tag{16}$$

式中：$F$——IES 低碳经济运行成本，美元；$f_{\mathrm{buy}}^{\mathrm{price}}$、$f_{\mathrm{CO_2}}^{\mathrm{price}}$、$f_{\mathrm{WI}}^{\mathrm{price}}$、$f_{\mathrm{PH}}^{\mathrm{price}}$、$f_{\mathrm{P2G}}^{\mathrm{price}}$、$f_{\mathrm{Q,cut}}^{\mathrm{price}}$、$f_{\mathrm{W}}^{\mathrm{price}}$、$f_{\mathrm{G}}^{\mathrm{price}}$——购电成本、阶梯式碳交易成本、垃圾焚烧电厂成本、CHP 和 GB 成本、P2G 成本、弃风弃光成本、风电和光伏机组运行维护成本、火电机组成本，美元。

1）购电成本 $f_{\mathrm{buy}}^{\mathrm{price}}$

$$f_{\mathrm{buy}}^{\mathrm{price}} = \sum_{t=1}^{T} \alpha_t P_{\mathrm{buy}}(t) \tag{17}$$

式中：$P_{\mathrm{buy}}(t)$——$t$ 时刻的购电量，MW；$\alpha_t$——$t$ 时刻的电价，美元/MW。

2）阶梯式碳排放交易成本 $f_{\mathrm{CO_2}}^{\mathrm{price}}$

其计算公式如式(14)所示。

3）垃圾焚烧电厂碳排放成本 $f_{\mathrm{WI}}^{\mathrm{price}}$

垃圾焚烧电厂运行无燃料成本，只需承担碳排放惩罚成本。

$$f_{\mathrm{WI}}^{\mathrm{price}} = P_{\mathrm{WI}} c_{\mathrm{w}} (e^{\alpha} - \gamma^{c}) \tag{18}$$

式中：$P_{\mathrm{WI}}$——垃圾焚烧发电功率，MW；$c_{\mathrm{w}}$——垃圾焚烧电厂参与碳交易系数；$e^{\alpha}$——单位电量碳排放量，取 0.528 t/MWh；$\gamma^{c}$——单位电量碳排放基准额，取 0.472 t/MWh。

4）CHP 和 GB 成本 $f_{\mathrm{PH}}^{\mathrm{price}}$

CHP 和 GB 的成本主要为燃料成本。

$$f_{\mathrm{PH}}^{\mathrm{price}} = k_{\mathrm{CH_4}} V_{\mathrm{buy}} \tag{19}$$

$$V_{\mathrm{buy}} = V_{\mathrm{CHP}} + V_{\mathrm{GB}} - V_{\mathrm{P2G}} \tag{20}$$

式中：$k_{\mathrm{CH_4}}$——单位天然气价格，美元/$\mathrm{m}^3$；$V_{\mathrm{buy}}$——天然气购买量，$\mathrm{m}^3$。

5）P2G 成本 $f_{\mathrm{P2G}}^{\mathrm{price}}$

P2G 的成本主要为原料成本和运行成本。

$$f_{\mathrm{P2G}}^{\mathrm{price}} = k_{\mathrm{CO_2}} Q_{\mathrm{buy}} + k_{\mathrm{P2G}} P_{\mathrm{P2G}} \tag{21}$$

$$Q_{\mathrm{buy}} = Q_{\mathrm{P2G,sum}} - Q_{\mathrm{P2G}} \tag{22}$$

式中：$k_{\mathrm{CO_2}}$——购买 $\mathrm{CO_2}$ 的价格，美元/t；$Q_{\mathrm{buy}}$——$\mathrm{CO_2}$ 购买量，t；$k_{\mathrm{P2G}}$——P2G 运行成本系数；$Q_{\mathrm{P2G,sum}}$——$\mathrm{CO_2}$ 总消耗量，t；$Q_{\mathrm{P2G}}$——P2G 获得的 $\mathrm{CO_2}$ 量，t。

6）弃风弃光成本 $f_{\mathrm{Q,cut}}^{\mathrm{price}}$

$$f_{\mathrm{Q,cut}}^{\mathrm{price}} = \delta_{\mathrm{DG}} \sum_{t=1}^{T} P_{\mathrm{DG,cut}}(t) + \delta_{\mathrm{PV}} \sum_{t=1}^{T} P_{\mathrm{PV,cut}}(t) \tag{23}$$

式中：$\delta_{\mathrm{DG}}$、$\delta_{\mathrm{PV}}$——单位弃风、弃光成本系数；$P_{\mathrm{DG,cut}}(t)$、

$P_{\mathrm{PV,cut}}(t)$——$t$ 时刻的弃风、弃光功率，MW。

7）风电和光伏机组运行维护成本 $f_{\mathrm{W}}^{\mathrm{price}}$

$$f_{\mathrm{W}}^{\mathrm{price}}=\lambda_1 P_{\mathrm{W}}+\lambda_2 P_{\mathrm{V}} \tag{24}$$

式中：$\lambda_1$、$\lambda_2$——风电、光伏机组的单位运行维护成本系数，美元/MW；$P_{\mathrm{W}}$、$P_{\mathrm{V}}$——风电、光伏机组的发电出力，MW。

8）火电机组成本 $f_{\mathrm{G}}^{\mathrm{price}}$

$$f_{\mathrm{G}}^{\mathrm{price}}=c_{\mathrm{G}}\sum_{t=1}^{T}P_{\mathrm{G}}(t) \tag{25}$$

式中：$c_{\mathrm{G}}$——$t$ 时刻机组运行的功率成本系数，美元/MW。

## 3.2 约束条件

1）风电、光伏出力约束

$$0\le P_{\mathrm{DG}}(t)\le P_{\mathrm{DG}}^{\max}(t) \tag{26}$$

$$0\le P_{\mathrm{PV}}(t)\le P_{\mathrm{PV}}^{\max}(t) \tag{27}$$

式中：$P_{\mathrm{DG}}(t)$、$P_{\mathrm{PV}}(t)$——$t$ 时刻风电、光伏功率，MW；$P_{\mathrm{DG}}^{\max}(t)$、$P_{\mathrm{PV}}^{\max}(t)$——$t$ 时刻风电、光伏功率最大值，MW。

3）EL、MR、HFC、CHP 运行约束

其运行约束如式（8）~式（11）所示。

4）GB 运行约束

$$\begin{cases}P_{\mathrm{GB,h}}(t)=\eta_{\mathrm{GB}}P_{\mathrm{g,GB}}(t)\\ P_{\mathrm{g,GB}}^{\min}\le P_{\mathrm{g,GB}}\le P_{\mathrm{g,GB}}^{\max}\\ \Delta P_{\mathrm{g,GB}}^{\min}\le P_{\mathrm{g,GB}}(t+1)-P_{\mathrm{g,GB}}(t)\le \Delta P_{\mathrm{g,GB}}^{\max}\end{cases} \tag{28}$$

$$P_{\mathrm{G}}^{\min}\le P_{\mathrm{G}}\le P_{\mathrm{G}}^{\max} \tag{29}$$

式中：$\eta_{\mathrm{GB}}$——$t$ 时段输入 GB 的效率；$\Delta P_{\mathrm{g,GB}}^{\max}$、$\Delta P_{\mathrm{g,GB}}^{\min}$——GB 的爬坡最大值和最小值，MW；$P_{\mathrm{g,GB}}^{\max}$、$P_{\mathrm{g,GB}}^{\min}$——输入气功率的最大值和最小值，MW。

5）火电机组运行约束

$$P_{\mathrm{G}}^{\min}\le P_{\mathrm{G}}\le P_{\mathrm{G}}^{\max} \tag{30}$$

式中：$P_{\mathrm{G}}^{\max}$、$P_{\mathrm{G}}^{\min}$——输入功率的最大值和最小值，MW。

6）电、热储能运行约束

$$\begin{cases}0\le P_t^{\mathrm{ESC}}\le P^{\mathrm{ESC,max}}\mu_t^{\mathrm{ESC}}\\ 0\le P_t^{\mathrm{ESD}}\le P^{\mathrm{ESD,max}}\mu_t^{\mathrm{ESD}}\\ 0\le \mu_t^{\mathrm{ESC}}+\mu_t^{\mathrm{ESD}}\le 1\\ S^{\mathrm{ES,\,min}}\le S_t^{\mathrm{ES}}\le S^{\mathrm{ES,max}}\\ S_0^{\mathrm{ES}}=S_{24}^{\mathrm{ES}}\end{cases} \tag{31}$$

式中：$P^{\mathrm{ESC,max}}$、$P^{\mathrm{ESD,max}}$——充、放电功率最大值，MW；$\mu_t^{\mathrm{ESC}}$、$\mu_t^{\mathrm{ESD}}$——$t$ 时段电、热储能装置是否充放电、热能，是为 1，否为 0；$S^{\mathrm{ES,max}}$、$S^{\mathrm{ES,min}}$——储电容量最大、最小值，MWh；$S_0^{\mathrm{ES}}$、$S_{24}^{\mathrm{ES}}$——电储能在一天的始、末值，MWh [22]。

热储能约束与电储能一致，不再赘述。

7）电、热功率平衡约束

$$\begin{aligned}P_{\mathrm{e,buy}}(t)=&P_{\mathrm{e,Load}}(t)+P_{\mathrm{e,EL}}(t)+P_{\mathrm{ES}}^{\mathrm{e}}(t)+P_{\mathrm{Cf}}(t)-P_{\mathrm{DG}}(t)-\\&P_{\mathrm{PV}}(t)-P_{\mathrm{WI}}(t)-P_{\mathrm{CHP,e}}(t)-P_{\mathrm{GB}}(t)-\\&P_{\mathrm{HFC,e}}(t)-P_{\mathrm{G}}(t),0\le P_{\mathrm{e,buy}}(t)\le P_{\mathrm{e,buy}}^{\max}\end{aligned} \tag{32}$$

$$P_{\mathrm{h,Load}}(t)+P_{\mathrm{ES}}^{\mathrm{h}}(t)=P_{\mathrm{HFC,h}}(t)+P_{\mathrm{MR}}(t)+P_{\mathrm{CHP,h}}(t)+P_{\mathrm{GB,h}}(t)+P_{\mathrm{Q_1}}(t) \tag{33}$$

式中：$P_{\mathrm{e,Load}}(t)$、$P_{\mathrm{h,Load}}(t)$——$t$ 时刻电、热负荷，MW；$P_{\mathrm{ES}}^{\mathrm{e}}(t)$、$P_{\mathrm{ES}}^{\mathrm{h}}(t)$——$t$ 时刻输入电、热储能的功率，MW；$P_{\mathrm{e,buy}}^{\max}$——各时刻购电功率的最大值，MW；$P_{\mathrm{Q_1}}(t)$——余热回收装置回收的余热量，MW。

8）气、氢功率平衡约束

$$\begin{aligned}P_{\mathrm{g,buy}}(t)=&P_{\mathrm{g,Load}}(t)+P_{\mathrm{ES}}^{\mathrm{g}}(t)+P_{\mathrm{g,CHP}}(t)+\\&P_{\mathrm{g,GB}}(t)-P_{\mathrm{MR,g}}(t),0\le P_{\mathrm{g,buy}}(t)\le P_{\mathrm{g,buy}}^{\max}\end{aligned} \tag{34}$$

$$P_{\mathrm{EL,H_2}}(t)=P_{\mathrm{H_2,MR}}(t)+P_{\mathrm{H_2,HFC}}(t)+P_{\mathrm{ES}}^{\mathrm{H_2}}(t) \tag{35}$$

式中：$P_{\mathrm{ES}}^{\mathrm{g}}(t)$——$t$ 时刻储存输入天然气的功率，MW；$P_{\mathrm{g,Load}}(t)$——$t$ 时刻气负荷，MW；$P_{\mathrm{g,buy}}^{\max}$——各时刻购气的最大值，MW；$P_{\mathrm{ES}}^{\mathrm{H_2}}(t)$——$t$ 时刻输入氢储的功率，MW。

# 4 算例分析

## 4.1 算例描述

为了验证本文所提优化模型的有效性，将一天 24 小时作为一个周期进行验证，系统中负荷与风电出力预测如图 5 所示，参照某地区典型垃圾焚烧电厂并在其后装 $CO_2$ 分离装置和含热泵的余热回收利用装置，焚烧炉每日燃烧垃圾发电量恒定为 2000 MWh，烟气净化及烟气分离装置耗电系数取 0.25，CHP 最大电、热出力为 150 MW，机组最大电、热爬坡功率为 20 MW，喷淋塔效率取 0.8，热泵能效系数参照实际情况取 4，单位天然气价取 40 美元/MWh。燃气机组的碳排放权配额 $\delta_{\mathrm{g}}$=0.386 t/MWh，外网购电等效于传统燃煤机组，碳排放权配额 $\delta_{\mathrm{e}}$=0.797 t/MWh，单位弃风成本 $\delta_{\mathrm{DG}}$=28 美元/MWh，碳交易基价 $c$=35 美元/t，碳交易价格长度 $d$ =300 t，分时电价模型、阶梯型碳交易模型和机组碳排放的其他参数见文献[23-24]，其他设备参数如表 1 所示。

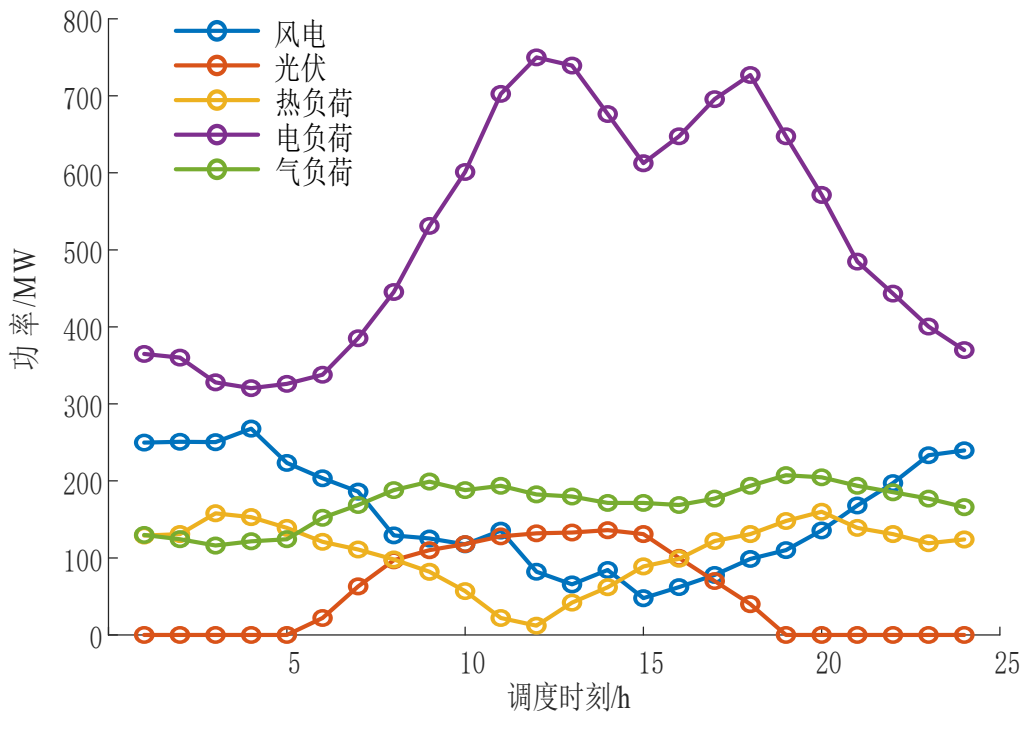

图 5 各负荷与风-光出力情况预测图

Fig. 5 Forecast diagram of each load and wind-photovoltaic power output

表 1 其他设备参数

Table 1　Equipment parameters

| 设备 | 容量/MW | 转换效率/% | 爬坡约束/% |
|---|---|---|---|
| EL | 100 | 87 | 20 |
| MR | 50 | 60 | 20 |
| HFC | 50 | 95 | 20 |
| GB | 60 | 95 | 20 |

## 4.2　不同运行模式构建及分析

为验证本文所述$CO_2$分离装置与含热泵的余热回收利用装置的垃圾焚烧电厂及细化电转气设备联合运行的优势，本文构建4种运行模式，均采用阶梯式碳交易机制，具体如表2所示。表3为4种运行模式下的成本结果分析。

表 2　4 种运行模式

Table 2　Four operating modes

| 项目 | 模式 1 | 模式 2 | 模式 3 | 模式 4 |
|---|---|---|---|---|
| 垃圾焚烧电厂-传统电转气 | √ | × | × | × |
| 垃圾焚烧电厂-细化电转气 | × | √ | √ | √ |
| 余热回收 | × | × | √ | √ |
| $CO_2$ 分离 | × | × | × | √ |
| 阶梯式碳交易 | √ | √ | √ | √ |

表 3　4 种运行模式成本对比

Table 3　Cost comparison of four operating modes

美元

| 成本 | 模式 1 | 模式 2 | 模式 3 | 模式 4 |
|---|---|---|---|---|
| 购电及购气 | 103291.86 | 76430.243 | 71518.554 | 69388.207 |
| 阶梯式碳交易 | 55949.638 | 52748.319 | 50385.049 | 41417.555 |
| P2G 运行 | 8899.0343 | 12904.042 | 11835.891 | 13330.929 |
| 弃风弃光 | 4216.373 | 2265.5666 | 1828.8848 | 845.3341 |
| 系统运行维护 | 95634.377 | 95634.377 | 95634.377 | 95634.377 |
| 火电机组 | 93654.906 | 92836.555 | 92409.392 | 92731.254 |
| 总计 | 361646.19 | 332819.1 | 323612.15 | 313347.66 |

1）模式 1 为垃圾焚烧电厂-传统 P2G 参与 IES 优化调度。如图 6 所示，在 21:00—24:00 和 00:00—07:00 阶段，电价与电负荷均处于低谷期，热负荷保持在高峰期，风电不断增加达到峰值，由于有大量的风力发电，在用电侧却没有足够的负荷与之平衡，且垃圾焚烧电厂及热电联产机组等有最小电出力和爬坡限制，由此产生了大量弃风。

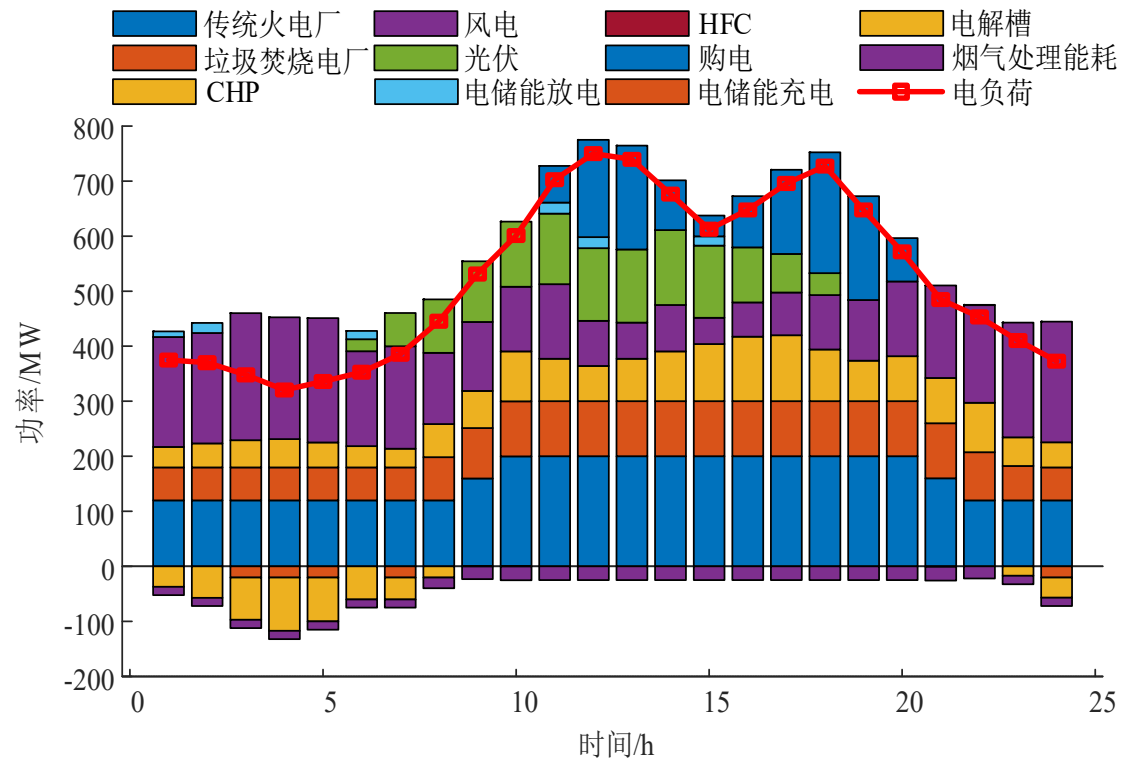

图 6　模式 1 电负荷和各供电单元出力

Fig. 6　Mode 1 electrical load and output of each power supply unit

2）图7所示为模式1和模式2的最优氢负荷图，模式2在模式1的基础上细化了电转气，利用P2G与EL设备，在21:00—24:00和00:00—07:00风电出力高的时段，将富余电能转化为氢能，在多数时段内氢能直接输送到HFC中转化为电、热能直接供给电、热负荷，消纳了弃风，且相比于传统的电转气先转化为天然气再经由燃气锅炉或热电联产机组燃烧供应，减少了一个能量转化环节，增加了能量转化效率。模式2相比于模式1的购电成本及购气成本大幅降低，减少了26861.6178美元，且HFC中氢能燃烧不产生二氧化碳，降低了外网购电、和燃气锅炉供热所产生的碳排放；而在00:00—06:00时，电负荷较小并伴有弃风现象，此时HFC产生的电能难以利用，氢气主要输送至MR合成天然气以及储氢罐中储存以达到消纳弃风的目的。模式2加装HFC后，P2G设备对整个系统的运行作用提高，也使得P2G运行成本提升。

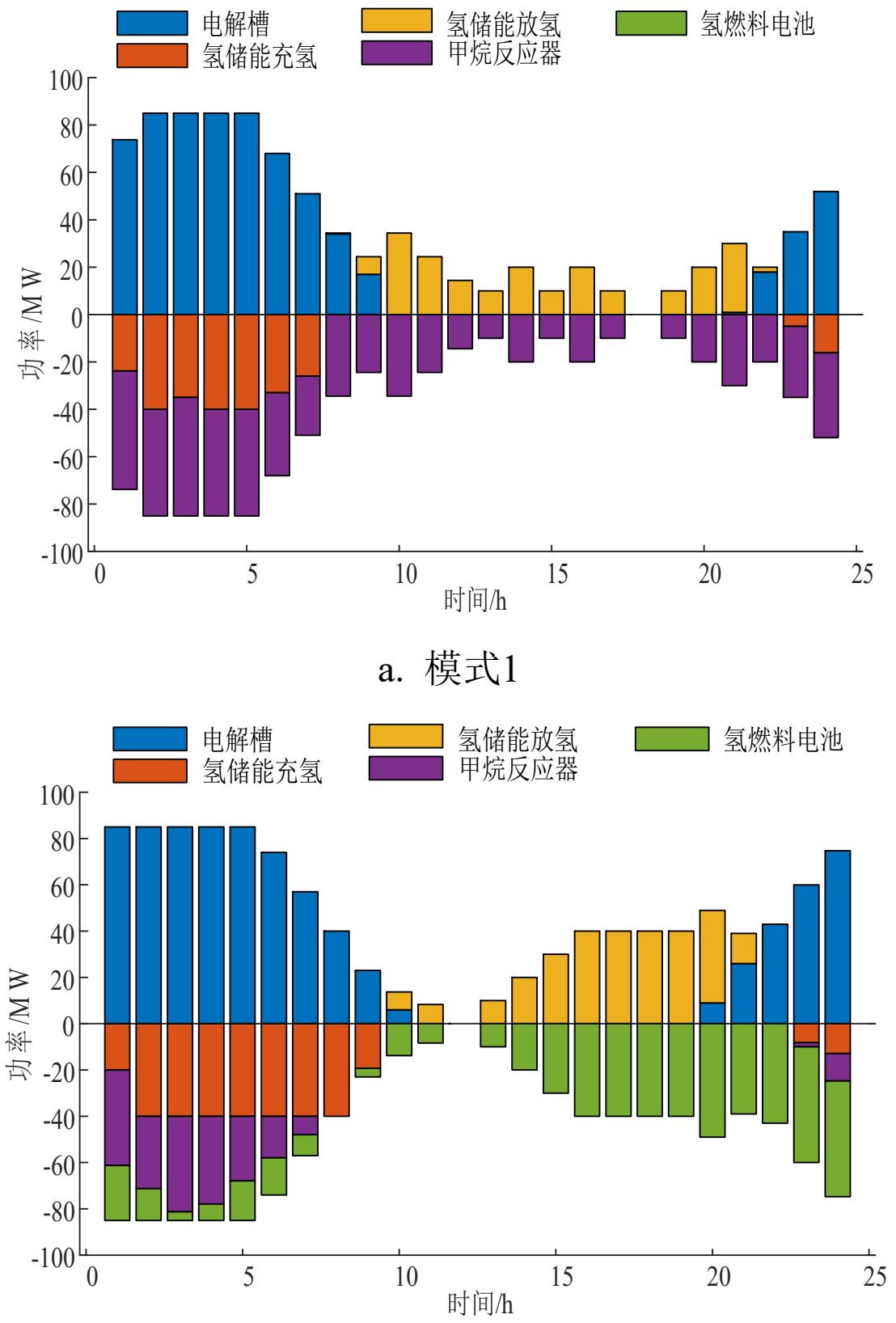

a. 模式1

b. 模式2

图 7　模式 1 和模式 2 的最优氢负荷

Fig. 7　Optimal hydrogen load for mode 1 and mode 2

3）图9所示为模式2和模式3的热负荷和各供热单元出力图。模式3在模式2的基础上增加了水源热泵、喷淋塔组成余热回收利用装置，考虑了甲烷化反应热能回收利用和垃圾焚烧电厂排出高温烟气的利用，在风电出力高峰时段，不仅电转气设备主动消纳风电，水源热泵在21:00—24:00和00:00—07:00非用电紧张时段通过消耗少量电能将高温烟气中的热量提取出来并提供给热网如图8所示，延伸了MR与EL消纳风电的价值，两者结合大大消纳了风电，弃风弃光成本相比于模式1减少了12387.4882美元。此运行模式一方面减少了GB以及CHP

热出力，另一方面削弱了CHP的热电比例耦合约束，使其在夜间用电高峰时增加电出力，减少购电量。余热回收利用装置与HFC配合参与调度，减少了HFC出力，降低了P2G运行成本，模式3相比于模式2的P2G运行成本减少了1068.1504美元。

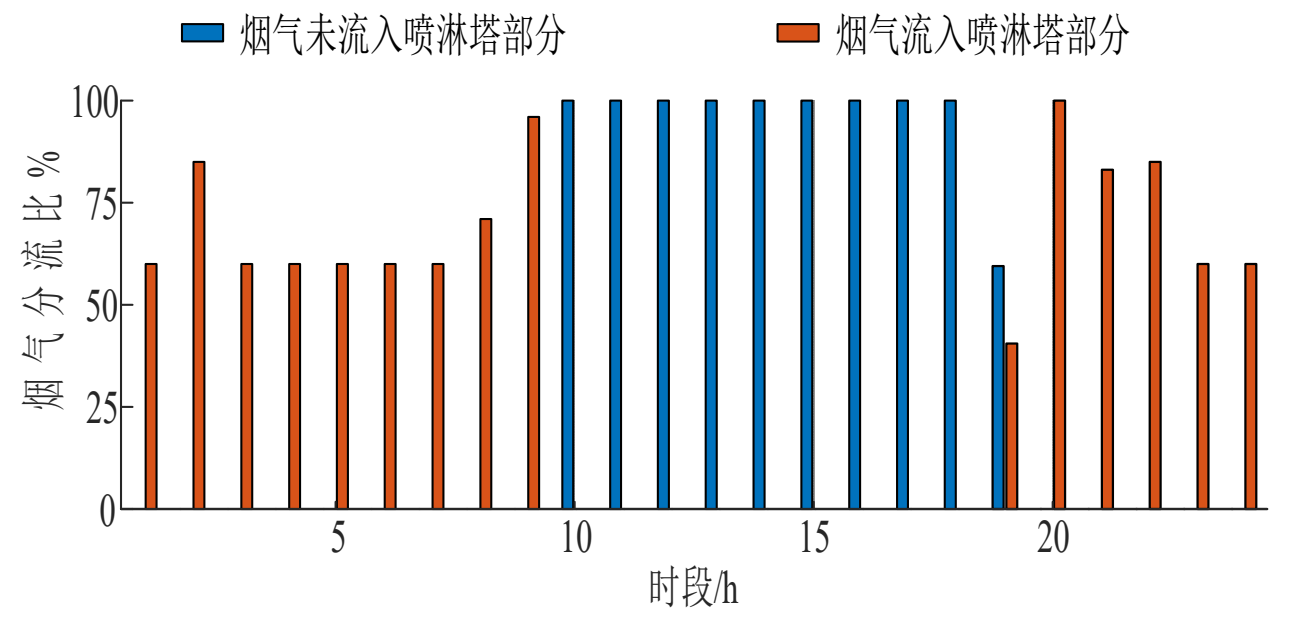

图8　烟气分流

Fig. 8　Smoke diversion

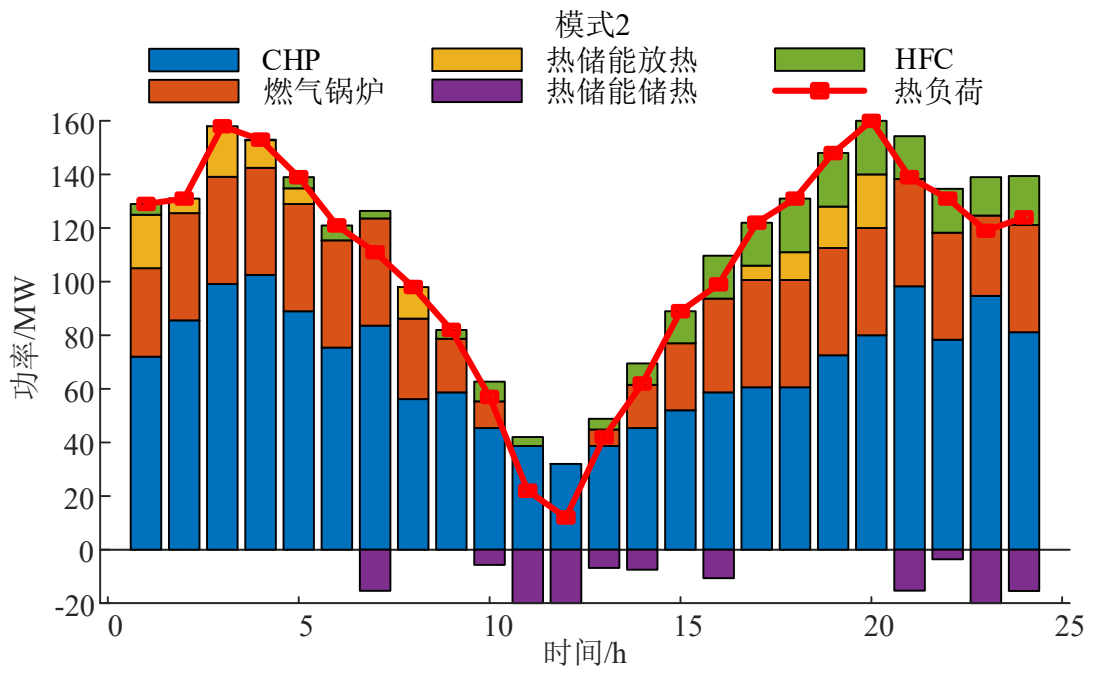

a. 模式2

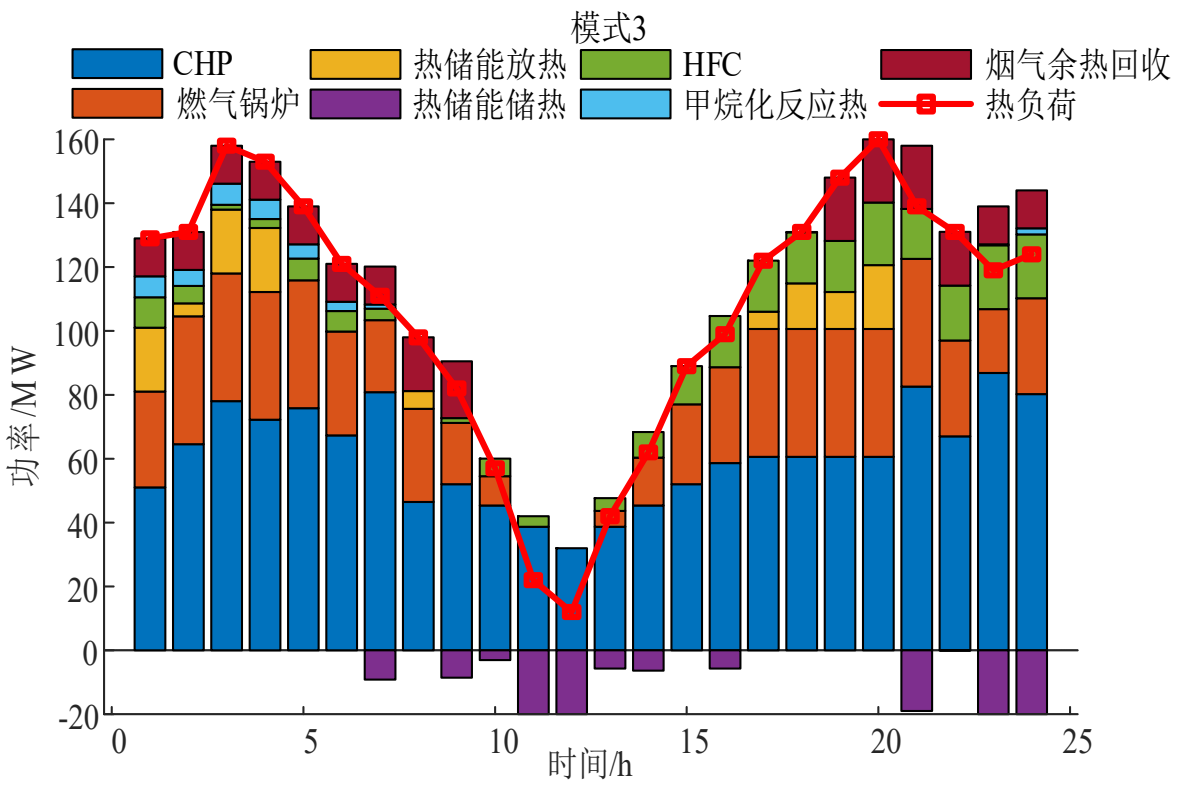

b. 模式

图9　模式2和模式3的热负荷和各供热单元出力

Fig. 9　Heating load and output of each heating unit in modes 2 and 3

4）图10所示为模式4下的电负荷和各供电单元出力图。模式4在模式3的基础上增加了$CO_2$分离装置，如图11所示在01:00—10:00和18:00—24:00电负荷较低以及风电较高时段灵活捕集$CO_2$，提高新能源消纳率，且分离出的$CO_2$可供给甲烷化反应装置，不仅降低碳排放，而且也降低了购碳成本，实现了碳循环，模式4的阶梯式碳交易成本大幅下降，相比于模式1减少了14532.0825美元。各运行模式下的弃风对比如图12所示。由图12可知，模式2考虑电转气的细化后可降低弃风率，模式3在此基础上考虑余热回收又可进一步降低弃风率，综合考虑各项因素的模式4下的弃风率最低。

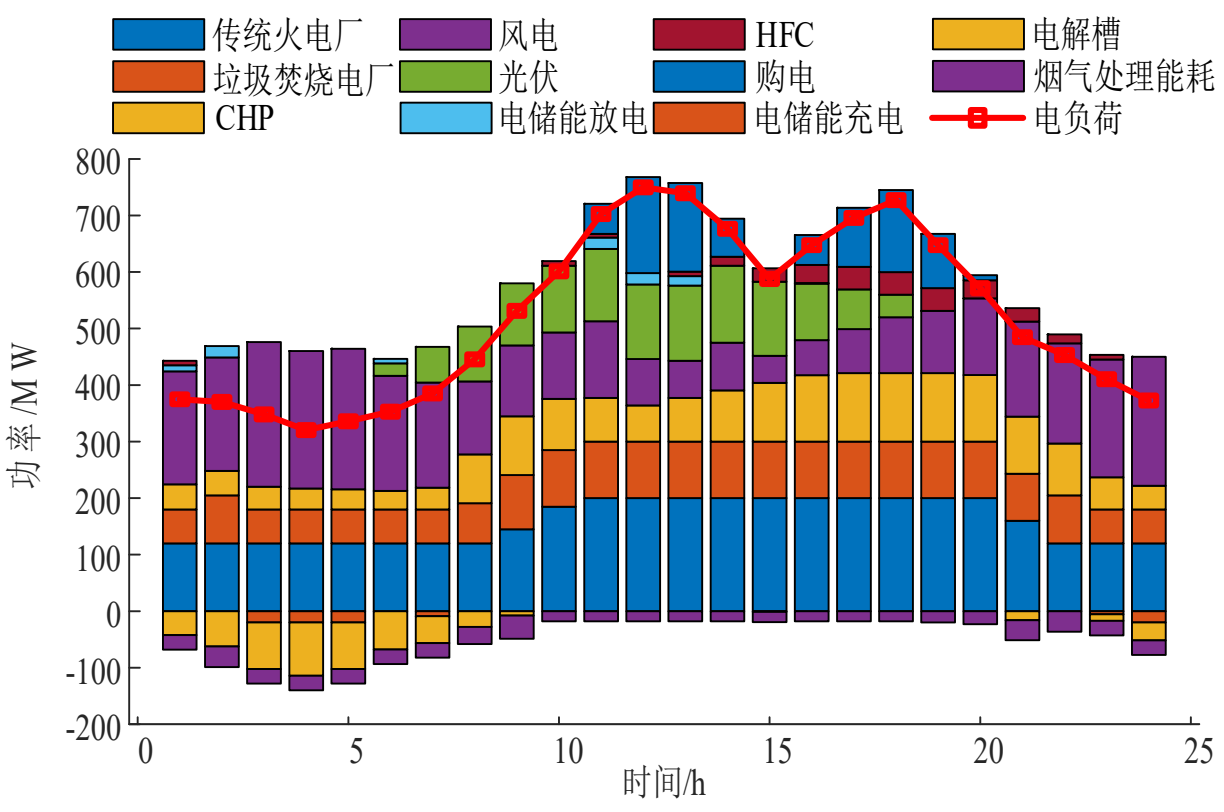

图10　模式4电负荷和各供电单元出力

Fig. 10　Mode 1 electrical load and output of each power supply unit

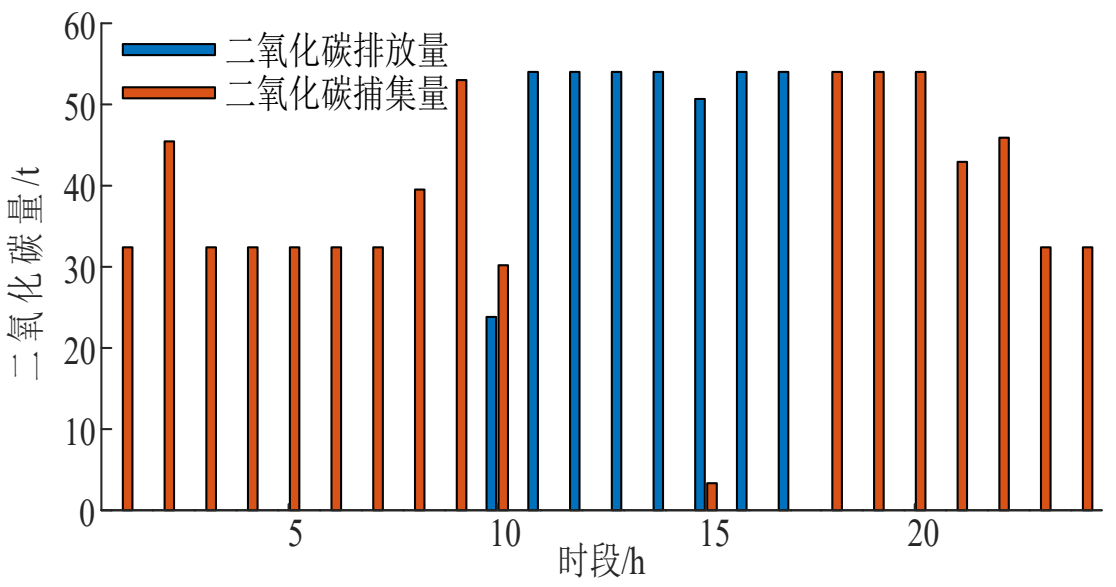

图11　$CO_2$排放与捕集量

Fig. 11　$CO_2$ emissions and capture capacity

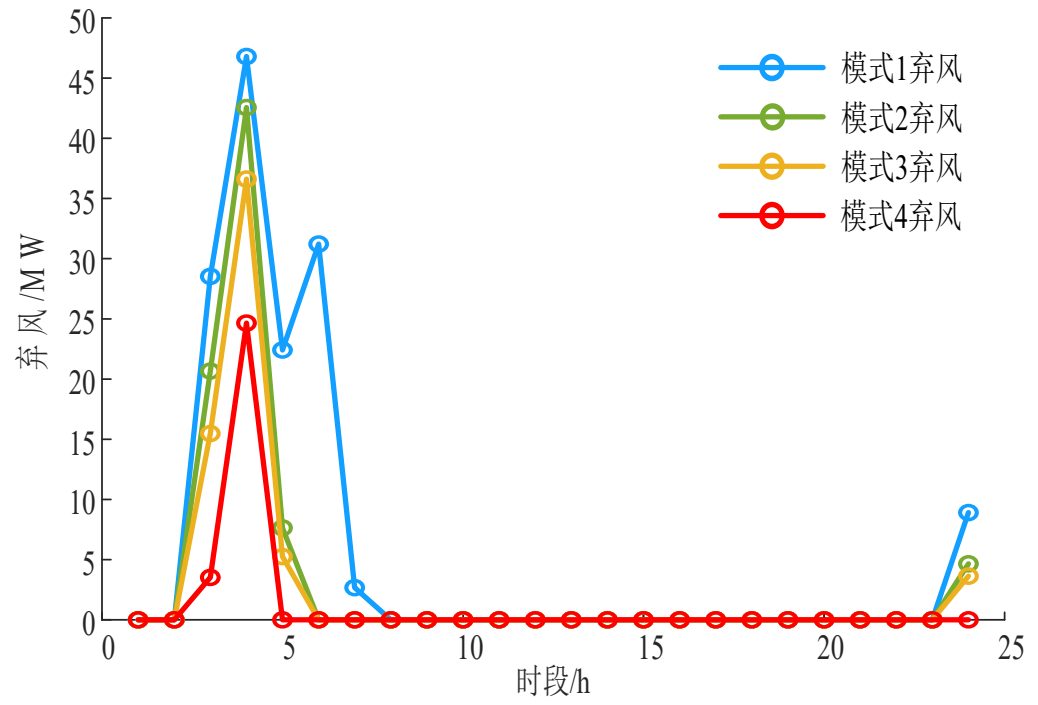

图12　4种运行模式下弃风对比

Fig. 12　Comparison of wind abandonment under four operating modes

综上，模式4实现了垃圾焚烧电厂-P2G协同运行下考虑烟气处理-余热回收的IES低碳优化调度。考虑的细化电制氢模型，可提高新能源和碳的利用率。烟气处理-余热回收进一步的限制了碳排放量以及实现了内部资源更为有效地利用，提升了平抑负荷波动的能力，达到削风填谷减少弃风弃光率的目的，模式4的总运行成本较模式1减少了48298.5326美元。

## 5　结　论

本文提出的垃圾焚烧电厂-P2G协同运行下考虑余热利用的IES低碳优化调度模型，引入了氢燃料电池、储

氢装置和考虑甲烷化反应余热利用以及细化电转气模型，将烟气处理-余热回收和$CO_2$分离装置进行联合运行，高效实现各单元之间电、热负荷的协调优化调度和碳的循环利用，提升了新能源消纳率和整体系统运行效益，通过研究分析，得出如下主要结论：

1）将传统P2G加入HFC装置，形成P2G两阶段运行过程，进一步促进了弃风的消纳，同时能够发挥氢能高能效的优势，减少能量的梯级损耗；并且HFC可承担一部分电、热的供能需求，减少18.7%的外网购电及购气成本并降低其产生的碳排放。

2）引入烟气处理-余热回收装置，考虑甲烷化反应释放热能的回收利用和垃圾焚烧电厂排出的高温烟气余热的利用，使用水源热泵装置，通过消耗少量电能将高温烟气中的热量提取出来并提供给热网，大大消纳风电，减少该部分热量所需购能成本6.7%，同时减少其他供热设备产生的碳排放。

3）考虑阶梯式碳交易成本，以提高清洁能源出力的同时减少碳排放。通过$CO_2$分离装置，使得垃圾焚烧电厂产生的$CO_2$被捕集起来，并输送至电转气中的甲烷化反应装置，实现了$CO_2$循环利用，减少了碳排放量，并降低了17.8%的碳交易成本，增加了系统的经济性。

[参考文献]

## LOW CARBON OPTIMAL SCHEDULING OF INTEGRATED ENERGY SYSTEM CONSIDERING WASTE HEAT UTILIZATION UNDER THE COORDINATED OPERATION OF INCINERATION POWER PLANT AND P2G

Wang Limeng, Wang Shuo, Wang Na, Ma Yuze, Li Yang

*(Key Laboratory of Modern Power System Simulation and Control & Renewable Energy Technology (Northeast Electric Power University), Ministry of Education, Jilin 132012, China)*

**Abstract:** In order to improve energy utilization and reduce carbon emissions, this paper presents a comprehensive energy system economic operation strategy of Incineration power plant Power-to-gas (P2G) with waste heat recovery. First, consider the coordinated operation of Incineration power plant - P2G, introduce the refined Power-to-gas two-stage operation process, add Hydrogen fuel cells on the basis of traditional Power-to-gas to reduce the energy ladder loss, and recycle the Methanation reaction heat; Secondly, in order to improve the energy utilization efficiency of Incineration, it is considered to install a waste heat recovery device containing a water source heat pump to recover the waste heat of flue gas and consume some electric energy, sourced from wind power, and add a $CO_2$ separation device to combine the recovered $CO_2$ with P2G to synthesize $CH_4$ to achieve carbon recycling. Finally, within the framework of a tiered carbon trading mechanism an IES optimization model for electricity-heat with the goal of minimizing the system operating cost is constructed, and the GUROBI modeling optimization engine is used to solve this model. The results verify the effectiveness of the model.